\documentclass[aps,amsmath,amssymb,groupedaddress,twocolumn]{revtex4}
\usepackage{graphicx}

\newcommand{\A}{{\text A}}
\newcommand{\B}{{\text B}}
\newcommand{\AB}{{\text{AB}}}

\begin{document}

\title{Entanglement entropy, black holes and holography}

\author{Roman~V.~Buniy} \email{roman@uoregon.edu}
\affiliation{Institute of Theoretical Science \\ University of Oregon,
Eugene, OR 97403}

\author{Stephen~D.~H.~Hsu} \email{hsu@duende.uoregon.edu}
\affiliation{Institute of Theoretical Science \\ University of Oregon,
Eugene, OR 97403}

\begin{abstract}
We observe that the entanglement entropy resulting from tracing over a
subregion of an initially pure state can grow faster than the surface
area of the subregion (indeed, proportional to the volume), in
contrast to examples studied previously. The pure states with this
property have long-range correlations between interior and exterior
modes and are constructed by purification of the desired density
matrix. We show that imposing a no-gravitational collapse condition on
the pure state is sufficient to exclude faster than area law entropy
scaling. This observation leads to an interpretation of holography as
an upper bound on the realizable entropy (entanglement or von Neumann) of a region,
rather than on the dimension of its Hilbert space.
\end{abstract}


\maketitle

\date{today}

\bigskip

\bigskip
Black holes radiate \cite{Hawking} and have entropy
\cite{Bekenstein}. The nature of this entropy is one of the great
mysteries of modern physics, especially due to its non-extensive
nature: it scales as the area of the black hole (in Planck units),
rather than its volume. This peculiar property has led to the holographic
conjecture \cite{HoltHooft,HolSusskind} proposing that the number of
degrees of freedom in any region of our universe grows only as the
area of its boundary. (See \cite{Bousso} for a review and discussion
of covariant generalizations of this idea.) The AdS/CFT correspondence
\cite{AdSCFTreview} is an explicit realization of holography.

The entropy of a thermodynamic system is the logarithm of the number
of the available microstates of the system, subject to some
macroscopic constraints such as fixed total energy. In certain string
theory black holes, these states have been counted explicitly
\cite{SV, MSW}. It has also been proposed that black hole entropy is
simply the entropy of quantum entanglement between the causally
disconnected interior and exterior of the hole. A description of black
hole radiance as originating from entanglement has been known for some
time \cite{Israel}. Starting with a pure state $\vert \psi
\rangle$, and tracing over a subspace (analogous to the black hole
volume), one obtains a density matrix with non-zero von Neumann
entropy (also known as entanglement entropy).  When $\vert \psi
\rangle$ is the ground state of a local quantum field theory (or, more
generally, any state with short range quantum correlations), the
resulting entropy also exhibits area scaling \cite{Sorkin, Srednicki,
EE, Ram,Plenio:2004he}. Recent work \cite{Einhorn} argues that even the
string theory microstate entropy can be attributed to entanglement.

In this note we observe that the entanglement entropy does not
necessarily scale as the area of the region which is traced
over. Indeed, it is easy to obtain pure states which lead to maximal
entropy, scaling as the volume. Such states have, as one might expect,
long range correlations between interior and exterior
modes. Entanglement entropy which scales like volume would seem to
contradict the usual area law results for black holes. We show that
this contradiction is evaded once gravitational effects are taken into
account: the maximal entropy states are subject to gravitational
collapse. If one requires that their construction not produce a black
hole larger than the original fiducial region, area bounds are
recovered. This line of reasoning suggests a new, less radical,
interpretation of holography which does not require the excision of
most of the states in Hilbert space. Instead, holographic bounds can
be interpreted as bounds on {\it realizable} entanglement entropy in
the presence of gravity.

We adopt the following notation in our discussion. The pure state
describing the entire system is $\vert \psi_{\AB} \rangle$. The
subregion over which we trace is B, and has volume $V$ and radius of
roughly $R$. We assume an ultraviolet regulator of order the Planck
scale and express dimensional quantities in Planck units.
 We take the Hilbert space ${\cal H}_\B$ to be that of the
QFT modes restricted to region B, so $\text{dim}({\cal H}_\B) = N =
c^V $ for some constant $c$. (For qubits, or individual spins on a lattice, $c=2$.) The exterior region is denoted by A and
has Hilbert space ${\cal H}_\A$.

We also remind the reader of the Schmidt decomposition theorem
\cite{QI}. Suppose $\vert \psi_{\AB} \rangle$ is a pure state of a
composite system $\AB$. Then there exist orthonormal states $\vert
\psi^{(n)}_\A \rangle$ for system A and $\vert \psi^{(n)}_\B \rangle$ for
system B such that
\begin{equation}
\label{SD}
\vert \psi_{\AB} \rangle = \sum_n \lambda_n^{\frac{1}{2}} \vert
\psi^{(n)}_\A \rangle \vert \psi^{(n)}_\B \rangle \, ,
\end{equation}
where $\lambda_n^\frac{1}{2}$ are nonnegative real numbers satisfying
$\sum_n \lambda_n = 1$. This is a simple consequence of the
singular value decomposition theorem. Note that the dimensionalities of ${\cal
H}_\A$ and ${\cal H}_\B$ might be very different, and that the range
over which the sum in Eq.~(\ref{SD}) runs is determined by the smaller
Hilbert space.

\bigskip

\noindent {\it Maximum entropy construction}

\bigskip

We first ignore gravitational effects and let B be an imaginary
subvolume (no black holes yet). We work backwards by first choosing
the density matrix $\rho_\B$ which maximizes entropy:
\begin{equation}
\rho_\B = N^{-1}\sum_{n=1}^N \vert \psi^{(n)}_\B \rangle \langle \psi^{(n)}_\B \vert.
\label{rhoB}
\end{equation} 
This results in entanglement entropy
\begin{equation}
S_\B = \ln N = V \ln c~,
\end{equation} 
which is maximal. The complementary density matrix, which describes
the mixed state resulting from tracing over the interior region B,
is
\begin{equation}
\rho_\A = N^{-1} \sum_{n=1}^N \vert \psi^{(n)}_\A \rangle \langle
\psi^{(n)}_\A \vert.
\label{rhoA}
\end{equation}
$\rho_\A$ has the same (non-zero) eigenvalues as $\rho_\B$, and hence
the same entropy, $S_A = S_B$.  The $\vert \psi^{(n)}_\A \rangle$ and
$\vert \psi^{(n)}_\B \rangle$ are orthogonal vectors forming the Schmidt basis.

We can construct a pure state 
\begin{equation} 
\label{pure}
\vert \psi_{\AB} \rangle = N^{-\frac{1}{2}} \sum_{n=1}^N \vert
\psi^{(n)}_\A \rangle \, \vert \psi^{(n)}_\B \rangle
\end{equation} 
which, upon tracing over B, yields the desired $\rho_\A$, and hence
volume behavior of the entropy. (This is possible for {\it any}
desired $\rho$, which can be seen from the Schmidt decomposition. The
procedure is known as ``purification'' of a mixed state.)
There are many such states, as an
$i$-dependent phase factor in the sum does not affect the resulting
density matrix. All such states exhibit substantial long range
correlations between interior and exterior modes.  This is easy to see, since there are only $c^A$ states localized near the boundary; correlations between $N \sim c^V$ interior and exterior states must have typical range at least as large as $R$. In fact, we expect
any states $\vert \psi_{\AB} \rangle$ with sufficiently large quantum
correlation lengths to yield entropies which scale like volume
\cite{Ram,Plenio:2004he}.

In the purification procedure, the Hilbert space ${\cal H}_\A$ is taken
isomorphic to ${\cal H}_\B$, but otherwise the physical nature of the
A states is unspecified. They could be QFT degrees of freedom in the
exterior or abstract qubits (spins) whose orientation states form
${\cal H}_\A$. We assume the latter case below, which means that the
A components of $\vert \psi_{\AB} \rangle$ can be manipulated (moved
about) without energetic consequences.

\bigskip

{\it Gedanken construction of black hole in region B}

\bigskip

Consider the consequences of an entanglement entropy $S$ which grows
like $V$. It seems reasonable to assume that the entanglement entropy
$S$ contributes to the Bekenstein-Hawking entropy
$S_{\text{BH}}$. In \cite{Einhorn} it is argued that $S_{\text{BH}} =
S$, although one might also imagine that $S_{\text{BH}} =
S + S',$ where $S'$ is an additional (positive) source of
entropy. Additional sources of entropy $S'$ might arise from coarse
graining---for example, characterizing the location of the horizon in
a classical way despite small Planck-length fluctuations in its
position.  However, these are likely to exhibit area scaling and do
not affect our arguments below.

Let the B region collapse into a black hole of volume $V$. Let the A
states be abstract qubits, which we spread thinly over the rest of the
universe (their correlations with the B states are independent of
their position, so are unaffected by the spreading). A semi-classical
calculation is then sufficient to determine the entropy of the black
hole. This yields the usual Bekenstein-Hawking entropy
$S_{\text{BH}}$, which scales like the area, in contradiction with an
entanglement entropy $S$ which scales as $V$.

\bigskip

{\it Holographic resolution}

\bigskip

The contradiction is resolved by noting that gravitational collapse
limits the number of states $N$ we can use in our construction of
$\vert \psi_{\AB} \rangle$. Note that this is subtly different from
limiting the actual size of ${\cal H}_\B$.

In the gedanken construction, it is reasonable to require that $\vert
\psi_{\AB} \rangle$ not have already undergone collapse to a black hole
larger than the region B.

Let the Hamiltonian be $H = H_\A + H_\B + H_{\AB}$. We take the A
states to be abstract qubits, and set $H_\A = 0$ by not allowing any
interactions between them. Since the desired state $\vert \psi_{\AB}
\rangle$ is a pure state, we can build it using unitary
transformations on any initial pure state. It is well-known in quantum
information theory that a general unitary transformation can be
efficiently implemented as the product of primitive manipulations,
each involving only a small number of qubits \cite{QI}. Therefore, in
each step of the construction of $\vert \psi_{\AB} \rangle$ the
interaction term $H_{\AB}$ need not be large and can neglected relative
to $H_\B$ for our purposes.

Then
\begin{equation} 
\langle \psi_{\AB} \vert H \vert \psi_{\AB} \rangle = N^{-1}
 \sum_{n=1}^N \langle \psi^{(n)}_\B \vert H_\B \vert \psi^{(n)}_\B \rangle
\end{equation} 
which is the average of expectation values $\langle H_\B \rangle$ in
states included in the $\vert \psi_{\AB} \rangle$ superposition.

The no-gravitational collapse requirement then implies (roughly) that
\cite{collapse}
\begin{equation}
\langle \psi_{\AB} \vert H_\B \vert \psi_{\AB} \rangle < R.
\end{equation}
Now, if one excludes states from the Hilbert space whose energies are
so large that they would have already caused gravitational collapse,
one obtains $\ln N < A^{\frac{3}{4}}$, as originally deduced by 't
Hooft \cite{HoltHooft}. 
't Hooft replaces the system under study with
a thermal one. The number of states of a system with constant total
energy $E$ is given to high accuracy by the thermal result in the
large volume limit (recall the microcanonical ensemble in statistical
mechanics). Given a thermal region of radius $R$ and temperature $T$,
we have $S_{\text{th}} \sim T^3 R^3$ and $E \sim T^4 R^3$. Requiring
$E < R$ then implies $S_{\text{th}} < R^{\frac{3}{2}} \sim
A^{\frac{3}{4}}$. These relations also imply
\begin{equation}
\label{dos}
E R \sim S_{\text{th}}^{\frac{4}{3}}  \sim ( \ln \nu )^{\frac{4}{3}} ,
\end{equation}
where $\nu$ is the multiplicity of states with total energy $E$. 
We stress that the
thermal replacement is just a calculational trick: temperature plays
no role in our results, and Eq.~(\ref{dos}) can be 
obtained also by direct counting.

Avoiding collapse to a black hole larger than $R$ requires that we cut
off the sum in (\ref{pure}) at of order $N \sim \exp A^{\frac{3}{4}}$,
well before the maximum $N = c^V$, which prevents our construction of
states $\vert \psi_{\AB} \rangle$ whose density matrices violate the
area bound.

This conclusion is more general than our original
construction. Previously we began with states of maximal entropy and
only subsequently imposed the no-gravitational collapse condition. In
the following we maximize the entropy subject to the collapse
condition (the two steps do not commute). The resulting density matrices are canonical ensembles,
with Boltzmann probabilities, in contrast to the equal probabilities
in (\ref{rhoB}) and (\ref{rhoA}). Nevertheless, the resulting upper
bound on entropy scales only as $A^{\frac{3}{4}}$.

Let $\vert \psi_{\AB} \rangle$ be an arbitrary pure state, and
consider the question of gravitational collapse in the region
B. Whether collapse occurs depends on local properties in B, so we can
trace over the A degrees of freedom and consider the resulting density
matrix
\begin{equation}
\label{med}
\rho_\B = \sum_{n=1}^N \lambda_n \vert \psi^{(n)}_\B \rangle \langle
\psi^{(n)}_\B \vert.
\end{equation} 
At this point the bound we derive on $S$ could either be interpreted as a bound on entanglement entropy, or simply a bound on the usual von Neumann entropy of the state which collapses to form the black hole. 
A reasonable no-collapse criteria is then
\begin{equation}
{\rm tr} ( \rho_\B H_\B ) = \langle H_\B \rangle < R.
\end{equation}
Adopting an energy eigenstate basis, we have the condition
\begin{equation}
\label{noc}
\sum_{n=1}^N \lambda_n E_\B^{(n)} < R,
\end{equation}
where the $\lambda_n$ (eigenvalues of $\rho_\B$) are the probabilities
for finding the system in energy eigenstate $n$. Note the requirement
that the gravitational field produced by the matter in region B is
semi-classical (so that the hoop conjecture or some similar collapse
criteria \cite{collapse} can be applied) may constrain the
distribution of $\lambda_n$ even more, requiring it to be highly
peaked around some central value. We do not impose this condition in
our analysis, though it likely strengthens our results.

Let us maximize 
\begin{eqnarray}
S=-\sum_n \lambda_n\ln{\lambda_n},\label{entropy}
\end{eqnarray}
where the $\{\lambda_n\}$ are subject to constraints $\sum_n
\lambda_n=1$ and
\begin{eqnarray}
\sum_n \lambda_n\epsilon_n = A,\label{no-collapse}
\end{eqnarray}
where $\epsilon_n = E_\B^{(n)} R$. We impose equality in
(\ref{no-collapse}) knowing that, since the density of states grows
with energy, the entropy will be maximized when the total energy of
the system is also maximal. Using the method of Lagrange multipliers
with
\begin{eqnarray}
\tilde S=-\sum_n \lambda_n\ln{\lambda_n} &+&\alpha(\sum_n
\lambda_n-1)\nonumber\\ &-& \beta(\sum_n \lambda_n\epsilon_n -A),
\end{eqnarray}
variations with respect to $\lambda_n$ and $\alpha$ give
\begin{eqnarray}
\label{lambdacanon}
\lambda_n=Z^{-1}e^{-\beta\epsilon_n}\label{lambda},
\end{eqnarray}
where
\begin{eqnarray}
\label{Z}
Z(\beta)=\sum_n e^{-\beta\epsilon_n}.
\end{eqnarray}
The resulting entropy,
\begin{eqnarray}
\tilde S(\beta)=\ln{Z}(\beta)-\beta A,
\end{eqnarray}
reaches its maximum at the point $\beta=\beta_*$, which is the
solution to the equation
\begin{eqnarray}
Z'(\beta)+AZ(\beta)=0.\label{beta}
\end{eqnarray}
It is easy to see that this equation always has one solution.
Finally, using Eqs.~(\ref{entropy}), (\ref{lambda}) and (\ref{beta}),
we find
\begin{eqnarray}
\max S=\ln{Z(\beta_*)}+\beta_* A.\label{entropy.max}
\end{eqnarray}

We now express the maximum entropy in terms of the density of
states. We start by replacing the sum in $Z(\beta)$ by the integral,
\begin{eqnarray}
Z(\beta)=\int d\epsilon\,\nu(\epsilon)e^{-\beta\epsilon},
\end{eqnarray}
where $\nu(\epsilon)=dn(\epsilon)/d\epsilon$, and evaluate it by the
saddle point method. The result is
\begin{eqnarray}
\ln{Z(\beta)}\approx \ln{\nu(\epsilon_*)-\beta\epsilon_*},
\end{eqnarray}
where $\epsilon=\epsilon_*(\beta)$ is the solution of
$d \ln \nu(\epsilon)/d\epsilon=\beta$. Eq.~(\ref{beta}) becomes
$\epsilon_*(\beta)\approx A$ with a solution $\beta=\beta_*(A)$. From
Eq.~(\ref{entropy.max}), the maximum entropy is simply
\begin{eqnarray}
\max
S\approx\ln{\nu(\epsilon)}\vert_{\epsilon=A}.\label{entropy.max.2}
\end{eqnarray}
From Eq.~(\ref{dos}), we immediately find
\begin{equation}
\max{S}\approx A^\frac{3}{4}.
\end{equation}
Since the maximally entropic system is thermal (see (\ref{Z})), and
the temperature $T = \beta^{-1}$ is determined by the condition
(\ref{no-collapse}) that the average total energy be $R$, our result
inevitably agrees with 't Hooft's calculation described earlier. One
easily generalizes to $d$ dimensions to obtain $A^{(d-1)/d}$ scaling.

Note, we have assumed simple boundary conditions (appropriate to a
finite box) in our calculation. While this is sufficient to count
states which might contribute extensively (i.e., as $V$) to the
entanglement entropy via long range correlations, it does not properly
treat contributions resulting from short range correlations at the
boundary of $B$, which are order $A$ \cite{Sorkin, Srednicki, EE,
Ram,Plenio:2004he,Einhorn}. It is interesting that once the gravitational
collapse condition is imposed, the maximal contribution of `bulk'
states (those not localized near the boundary) is reduced from volume
scaling to $A^{\frac{3}{4}}$, which is smaller than the original $A$
scaling from boundary correlations.

\bigskip
{\it Conclusions}
\bigskip

The holographic conjecture makes the rather strong assertion that
states with $\langle i \vert H \vert i \rangle$ greater than $R$
simply {\it do not exist} in the Hilbert space. This conjecture,
however, leads to a number of puzzles. It is unclear what becomes of
locality or how unitarity is preserved. Do the high energy states
participate in virtual processes? How is the path integral measure
modified? Of course the biggest puzzle related to holography is why
the universe appears to have $d$ spacetime dimensions if the Hilbert
space is only that of a $d-1$ dimensional system.

We suggest an alternative interpretation of black hole entropy bounds.
The gravitational collapse condition on $\vert \psi_{\AB} \rangle$
places an upper bound on the entanglement (or von Neumann) entropy that can be realized
from a region B without forming a black hole larger than B
itself. Highly energetic states remain in the theory (they appear explicitly in the maximal entropy states; see Eq.~(\ref{med}) and (\ref{lambdacanon})), but cannot increase the entanglement entropy beyond the area of B in
Planck units. Entropy bounds reflect the limitations that gravity
imposes on the construction of pure states or density matrices, but do
not require a truncation of the Hilbert space itself.

\bigskip
\emph{Acknowledgements.---} The authors thank A. Zee for
useful comments. The authors are supported by the Department of Energy
under DE-FG06-85ER40224.


\bigskip


\end{document}